\documentclass[
    ,final            
  ]
  {aipproc}

\layoutstyle{8x11single}

\usepackage{array,amsmath}

\begin{document}

\title{Probing short-lived fluctuations in hadrons and nuclei}
\author{St\'ephane Munier}{address={Centre de physique th\'eorique, 
\'Ecole Polytechnique, CNRS, Palaiseau, France}
}
\begin{abstract}
We develop a picture of dipole-nucleus (namely dilute-dense) 
and dipole-dipole (dilute-dilute)
scattering in the high-energy regime
based on the analysis of the fluctuations
in the quantum evolution.
We emphasize the difference in the
nature of the fluctuations probed in these two processes
respectively, which, interestingly enough, leads to
observable differences in the scattering amplitude profiles.
\end{abstract}

\maketitle


This paper introduces and summarizes the results recently published
in Ref.~\cite{Mueller:2014fba}, from a less technical viewpoint 
(see Ref.~\cite{Munier:2014hga} for a complementary presentation), 
and with illustrations from numerical simulations 
(see Figs.~\ref{fig:dipole} and~\ref{fig:nucleus} below).
Our goal is to understand
the qualitative properties of the short-lived 
and short-distance (with respect to $1/\Lambda_\text{QCD}$)
quantum fluctuations, namely the ones that are probed most efficiently
in deep-inelastic scattering experiments in the small-$x_\text{Bj}$ regime,
or in observables in proton-proton and proton-nucleus
scattering whose cross sections
may be related to dipole amplitudes (see e.g. Ref.~\cite{Mueller:2012bn}).
(A recent 
general review of high-energy QCD can be 
found in Ref.~\cite{kovchegov2012quantum}).

We shall first describe qualitatively the scattering
of two color dipoles and of a dipole off a nucleus, before turning
to an analysis of the quantum fluctuations.
We shall eventually review the most striking quantitative prediction
derived from our discussion.

Our picture relies on the well-known 
color dipole model~\cite{Mueller:1993rr}, which
describes, in the framework of perturbative quantum
chromodynamics, how the quantum state of a hadron
builds up from a cascade of dipole splittings.

\section{Picture of the interaction of a small dipole with QCD matter}

\subsection{Scattering of a dipole off a dilute target and off a dense target}

We start with the scattering of
two color dipoles (concretely, e.g. two quark-antiquark pairs)
of respective transverse sizes $r_0$ and $R_0$
with the ordering $|r_0|<|R_0|$.

At low energy, the forward elastic scattering amplitude of
the dipoles consists in the exchange of a pair of gluons.
Since the dipoles are colorless, this exchange can take place only
if their sizes are comparable (on a logarithmic scale), 
and if the scattering occurs at 
coinciding impact parameters. Once these conditions are
fulfilled, the cross section is parametrically
proportional to~$\alpha_s^2$.

\begin{figure}[t]
\begin{tabular}{m{0.35\textwidth} | m{0.35\textwidth}}
  \centerline{\includegraphics[height=.15\textheight]{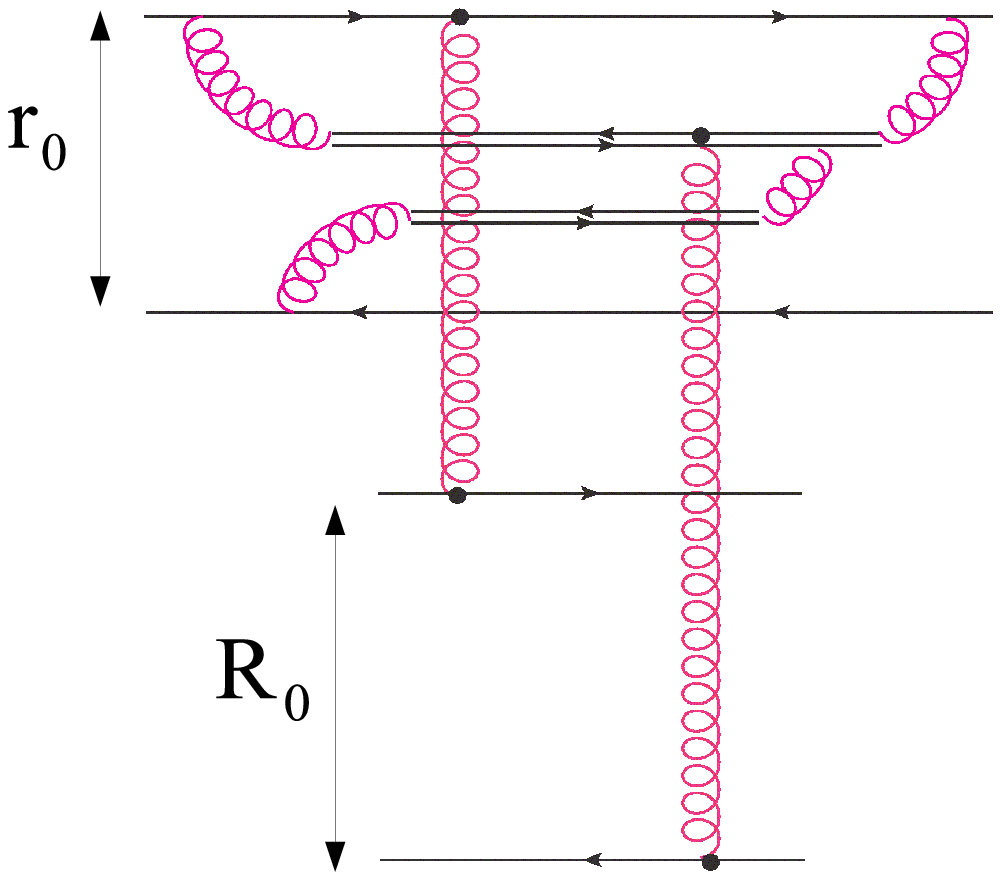}}
& \centerline{\includegraphics[height=.15\textheight]{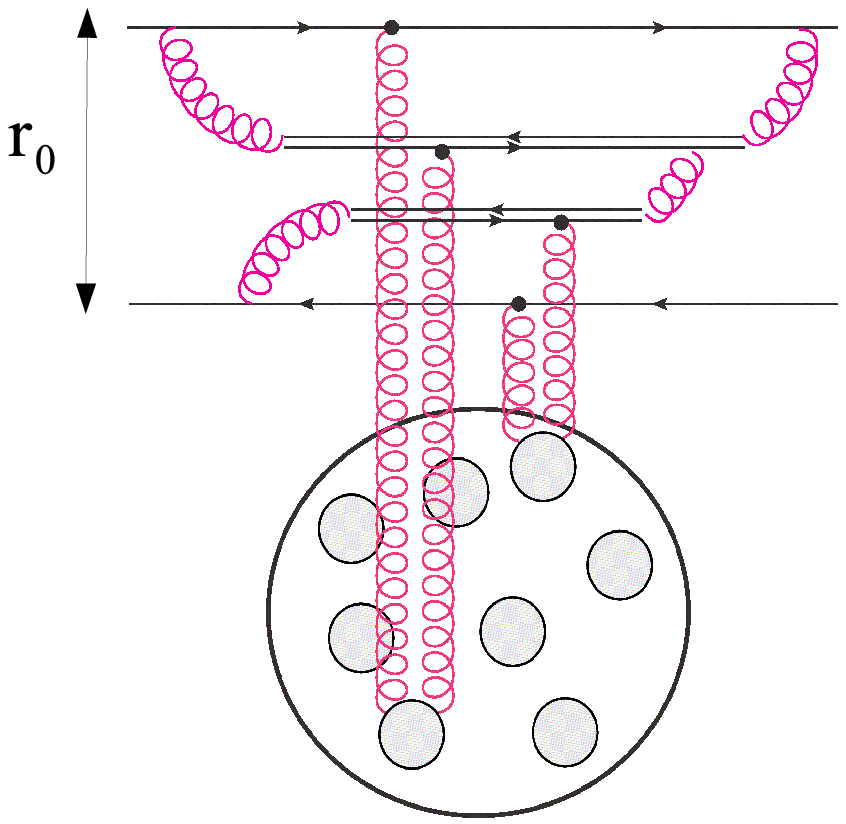}}\\
\multicolumn{1}{c}{(a)} & \multicolumn{1}{c}{(b)}
\end{tabular}
\caption{\label{fig:picture}
Diagrams contributing to the
forward elastic scattering amplitude of a dipole
off another dipole (a), and off
a nucleus (b). 
In the latter case, the nucleons, represented by
small disks,
are assumed independent and present in large numbers. 
This picture is valid in the restframe of the 
target dipole and nucleus, in such a way that all quantum evolution 
takes place in the
projectile dipole (initially of transverse size $r_0$). 
In the large number-of-color
limit, the Fock state of the 
projectile is a set of color
dipoles at the time of the interaction~\cite{Mueller:1993rr}.
In the dipole-dipole case, the scattering consists in 
the exchange of a pair of
gluons between one of the dipoles resulting from the evolution,
and the target dipole. In the dipole-nucleus case,
the dipoles whose size is larger than the inverse
saturation momentum of the nucleus exchange an arbitrary number
of pairs of gluons with
independent nucleons in the 
nucleus, eventually leading to a resummation {\it \`a la} Glauber-Mueller,
and physically, to absorption.}
\end{figure}

Let us go to larger center-of-mass 
energies $\sqrt{s}$
by boosting the small dipole to the 
rapidity $y=\ln (s R_0^2)$.
The most probable Fock state at the time
of the interaction is then a dense state of gluons, which
may be represented by a set of dipoles \cite{Mueller:1993rr}
(see the sketch in Fig.~\ref{fig:picture}a).
The amplitude is now enhanced by the number of
these dipoles which have a size of the order of $R_0$.
We may define a ``one-event amplitude'' $T_d^{\text{one event}}$, 
which is related to
the probability that gluons are exchanged between one given realization
of the dipole evolution and the target dipole.
If $n(r_0,r|y)$ denotes the number of dipoles in a given realization
of the evolution up to rapidity $y$ of size $r$, starting with
a dipole of size $r_0$, then
$T_d^{\text{one event}}(r_0,R_0|y)\simeq \alpha_s^2\times n(r_0,R_0|y)$,
where it is understood that the impact parameters 
of the dipoles which scatter
need to coincide
(up to a distance of order of the smallest size).
Of course, the dipole number fluctuates from event to event, 
and so does
$T_d^{\text{one event}}$.
The physical amplitude measured in an experiment is proportional to
the average of the latter over events, namely over
realizations of the dipole evolution.
We conclude that {\it the scattering of two dipoles of respective sizes
$r_0$ and $R_0$ probes the
density of gluons of transverse size $R_0$, at a given
impact parameter, in typical quantum
fluctuations of a source dipole of size $r_0$
appearing in the quantum evolution over the rapidity $y$}.\\

We turn to the case in which instead of
the dipole of size $R_0$, the target consists in a large
nucleus.
At low energy, the dipole now scatters through multiple
gluon exchanges since the nucleus is dense. 
The Glauber-Mueller summation of the latter
implemented in the McLerran-Venugopalan model~\cite{McLerran:1993ni}
gives the scattering amplitude in the form
$T_A(r_0)=1-e^{-r_0^2 Q_A^2/4}$.
A new momentum scale $Q_A$ is generated: It is 
the saturation momentum of the nucleus, and grows
with the nucleus mass number $A$ like $Q_A^2\sim A^{1/3}$.
This amplitude may actually be 
approximated
by a Heaviside $\Theta$ function:\footnote{This statement relies on the fact
that the exponential in the McLerran-Venugopalan formula is ``steep'', in some sense.
The full argument specifies what ``steep'' means, which is not completely
straightforward, see the discussion in Ref.~\cite{Mueller:2014fba}.}
$T_A(r_0)\simeq\Theta\left[\ln (r_0^2 Q_A^2/4)\right]$.
This has the advantage of simplifying the
interpretation of the scattering of the evolved dipole 
with the nucleus.

If one increases the center-of-mass energy by boosting
the dipole, then at the time of the interaction,
in each event, the nucleus ``sees'' a set of dipoles.
The interaction takes place if and only if at least one dipole
in this set
has a size larger than the inverse saturation momentum.
Hence
$T_A^{\text{one event}}(r_0,Q_A|y)\simeq 1$ if at least
one of the dipoles in the evolution is larger than 
$\sim 1/Q_A$, or
0 else.
The cross section that is measured in an experiment
by counting events 
(which technically is an averaging of $T_A$ over the events)
is thus the probability that at least one dipole
in the Fock state of the source dipole evolved to the time
of the interaction has a size larger than $\sim 1/Q_A$
(see Fig.~\ref{fig:picture}b).
Hence, while dipole-dipole scattering
probes the bulk of the quantum state, 
{\it the scattering of a dipole of size $r_0$ 
with a nucleus is sensitive to the probability distribution
of the size of the largest fluctuation generated 
in the quantum evolution of the dipole,}
namely, in a mathematical language, to the 
{\it statistics of extremes.}

\subsection{Rapidity evolution of a dipole}

We have just argued that scattering amplitudes can be thought of
as probes of the quantum fluctuations of the dipole.
The latter are best pictured with the help of the
color dipole model.
We shall now describe the way how these fluctuations build up 
(see Ref.~\cite{Mueller:2014gpa} for an extensive discussion), in order
to arrive at a model that leads to quantitative predictions for the
scattering amplitudes.

In these matters, useful intuition can be gained from numerical
implementations of the dipole model in the
form of a Monte Carlo event generator~\cite{Salam:1996nb}.
Therefore, we shall illustrate our arguments with the help
of simulations of the toy model constructed
in Ref.~\cite{Munier:2008cg}.\\

\begin{figure}[h]
\begin{tabular}{m{0.5\textwidth} | m{0.5\textwidth}}
  \includegraphics[width=.5\textwidth]{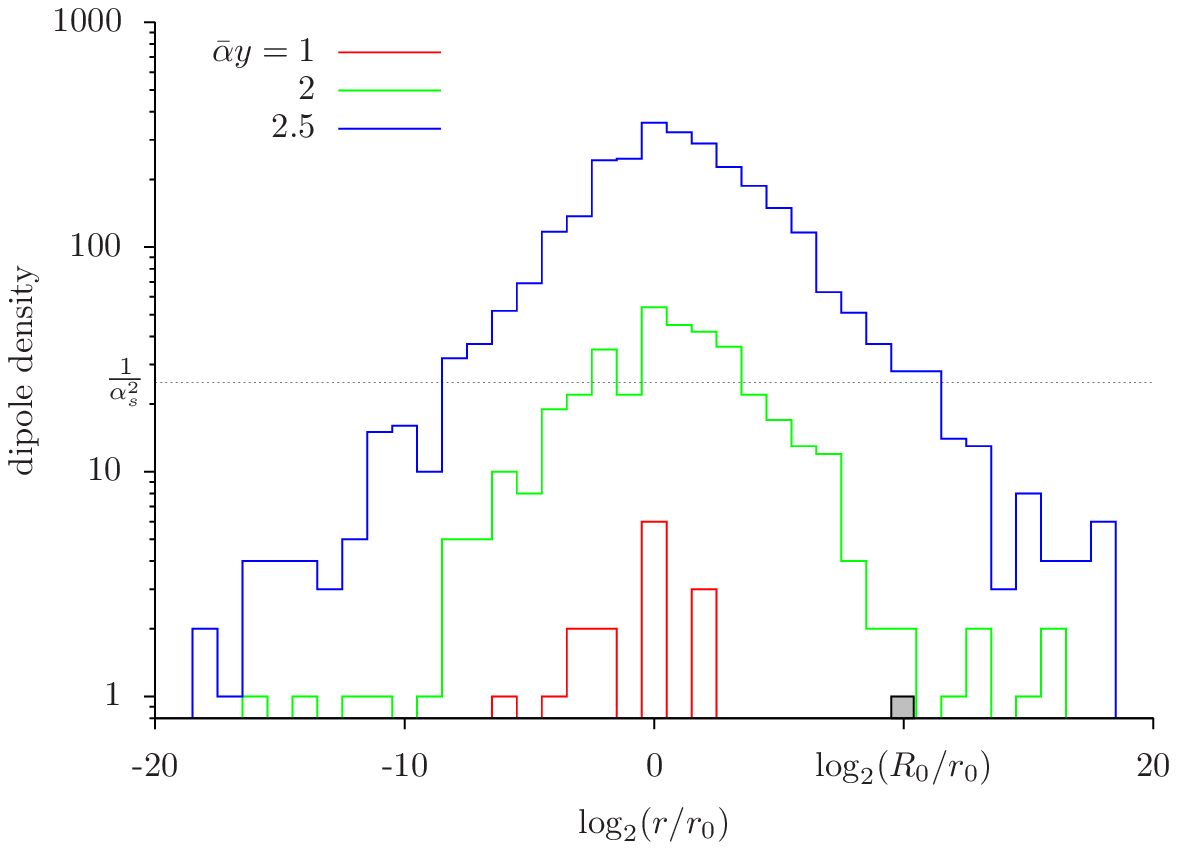}
& \includegraphics[width=.5\textwidth]{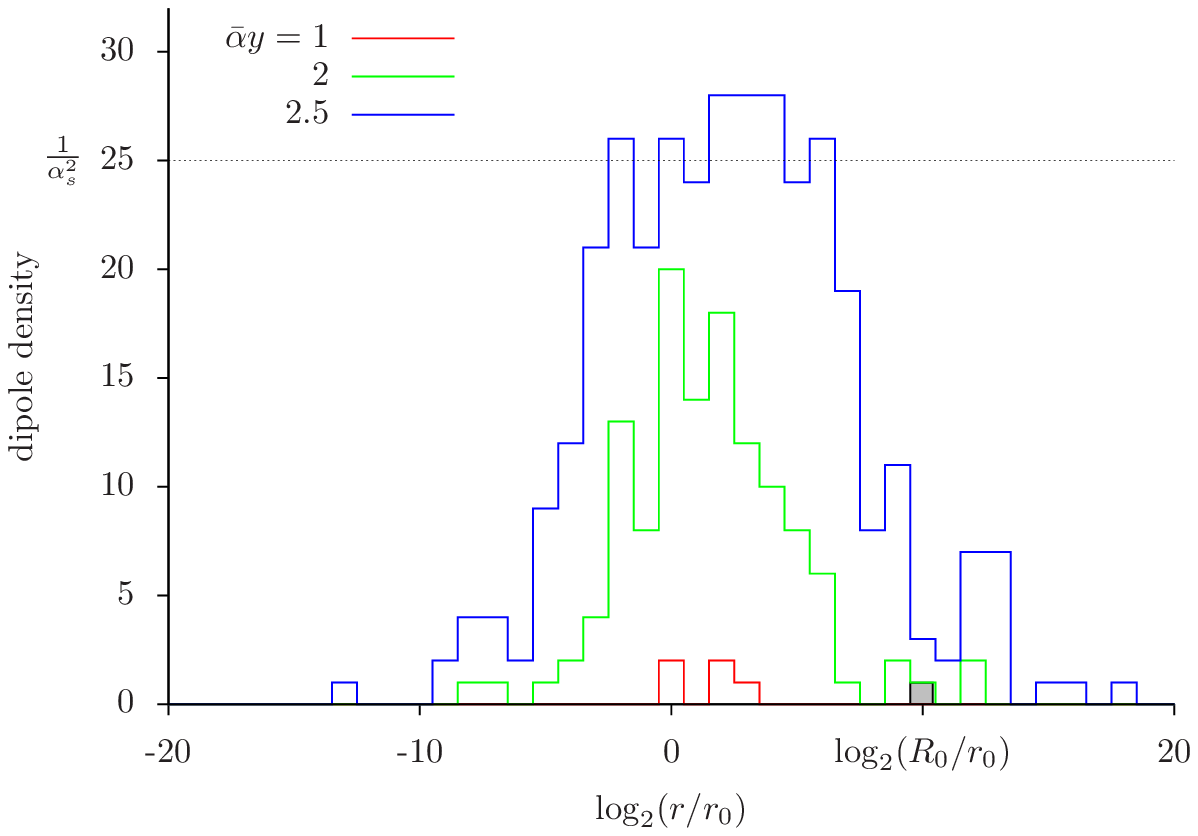}\\
\multicolumn{1}{c}{(a)} & \multicolumn{1}{c}{(b)}
\end{tabular}
\caption{Realizations of the dipole evolution in
the simplified model of Ref.~\cite{Munier:2008cg} for
three different rapidities. What is plotted is the
dipole number $n(r_0,r|y)$ {\it at zero impact parameter} in bins of
the logarithm of the dipole size (normalized to the size $r_0$ of
the dipole in the initial state).
The shaded rectangle of height unity 
indicates the target dipole of size $R_0$.
(a) No saturation in the dipole evolution. 
One sees that setting e.g. $\alpha_s=0.2$, unitarity
would be violated for $\bar\alpha y=2.5$ in this
event. (Note the logarithmic
scale on the $y$-axis).
(b) Saturation in the dipole evolution.
The maximum density of dipoles is around $1/\alpha_s^2=25$.
In both cases, one sees that the density of dipoles has fluctuations, which
self-average in the bins in which the density is large.
\label{fig:dipole}
}
\end{figure}

The quantum evolution of the dipole proceeds through successive
independent splittings $1\rightarrow 2$, with rate per unit rapidity
given by the probability
distribution
$\frac{dP}{dy}=\bar\alpha
\frac{r_0^2}{r_1^2(r_0-r_1)^2}
\frac{d^2 r_1}{2\pi}$,
where $r_0$ is the size vector of the initial dipole,
$r_1$ and $r_0-r_1$ are the size vectors of the offspring, and
$\bar\alpha\equiv \alpha_s N_c/\pi$.
The stochastic process of dipole evolution is a {\it branching diffusion}.
Indeed, if one looks at the distribution 
of the log of the dipole sizes
at a {\it given} 
impact parameter,
it has a Gaussian-like shape growing exponentially with the rapidity, 
see Fig.~\ref{fig:dipole}a.
The stochasticity has a strong effect wherever the number of dipoles is
of order unity, namely either in the beginning of the evolution (for low rapidities),
or at the edges of the distribution. Elsewhere, in the bulk of the
distribution, the law of large numbers applies, and therefore, the 
rapidity evolution of the 
dipole density is nearly deterministic.

According to these observations, 
in Ref.~\cite{Mueller:2014fba}, we conjectured that there are
essentially two types of fluctuations: The ``front fluctuations'' that
come from the early stages of the evolution,
and the ``tip fluctuations'' occurring throughout the
evolution in the regions in which the dipole density is low. 
Both kinds of fluctuations shift the distribution
of dipoles towards larger sizes,
by say $\Delta$ and $\delta$ respectively, in logarithmic scale.
We conjectured that the latter random variables are distributed exponentially,
$p(\Delta)\sim e^{-\gamma_0 \Delta}$ and $p(\delta)\sim e^{-\gamma_0\delta}$,
where $\gamma_0\simeq 0.63$ is a number determined by the eigenvalues of
the BFKL kernel.\\

\begin{figure}[h]
\begin{tabular}{m{0.5\textwidth} | m{0.5\textwidth}}
  \includegraphics[width=.5\textwidth]{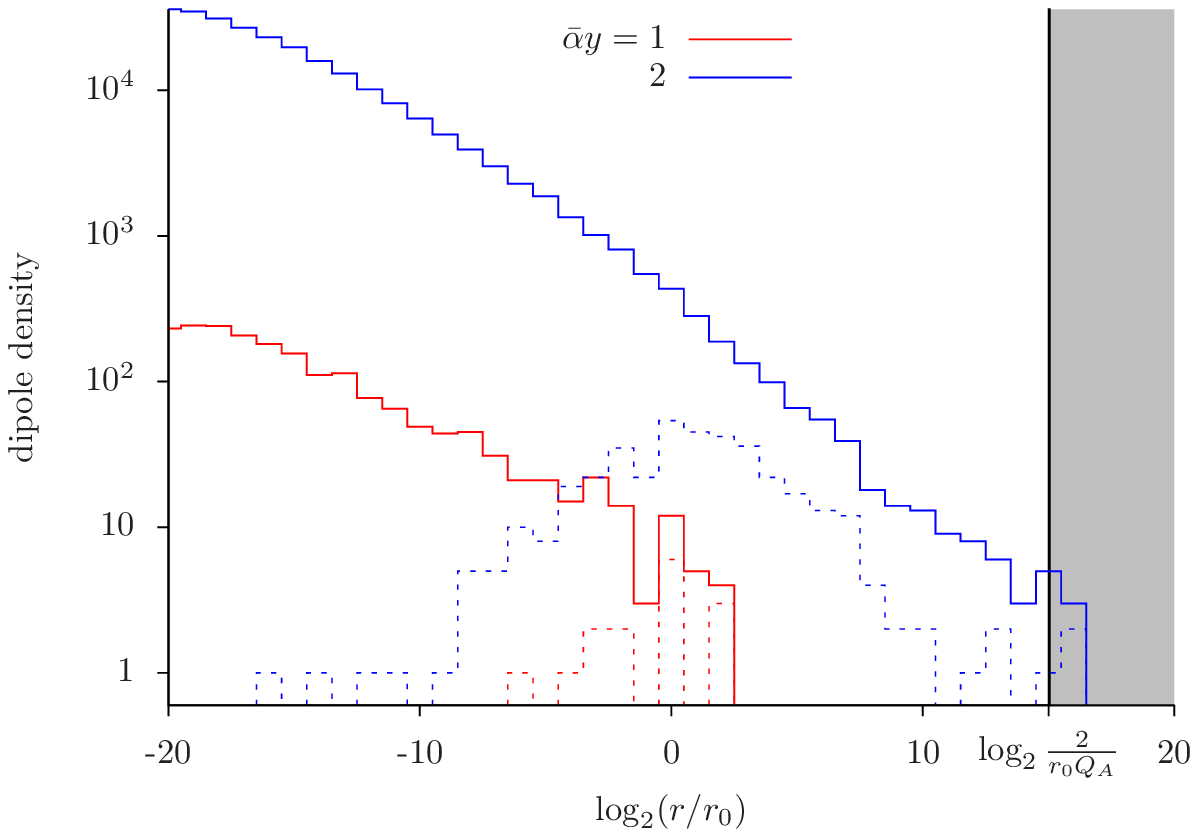}
& \includegraphics[width=.5\textwidth]{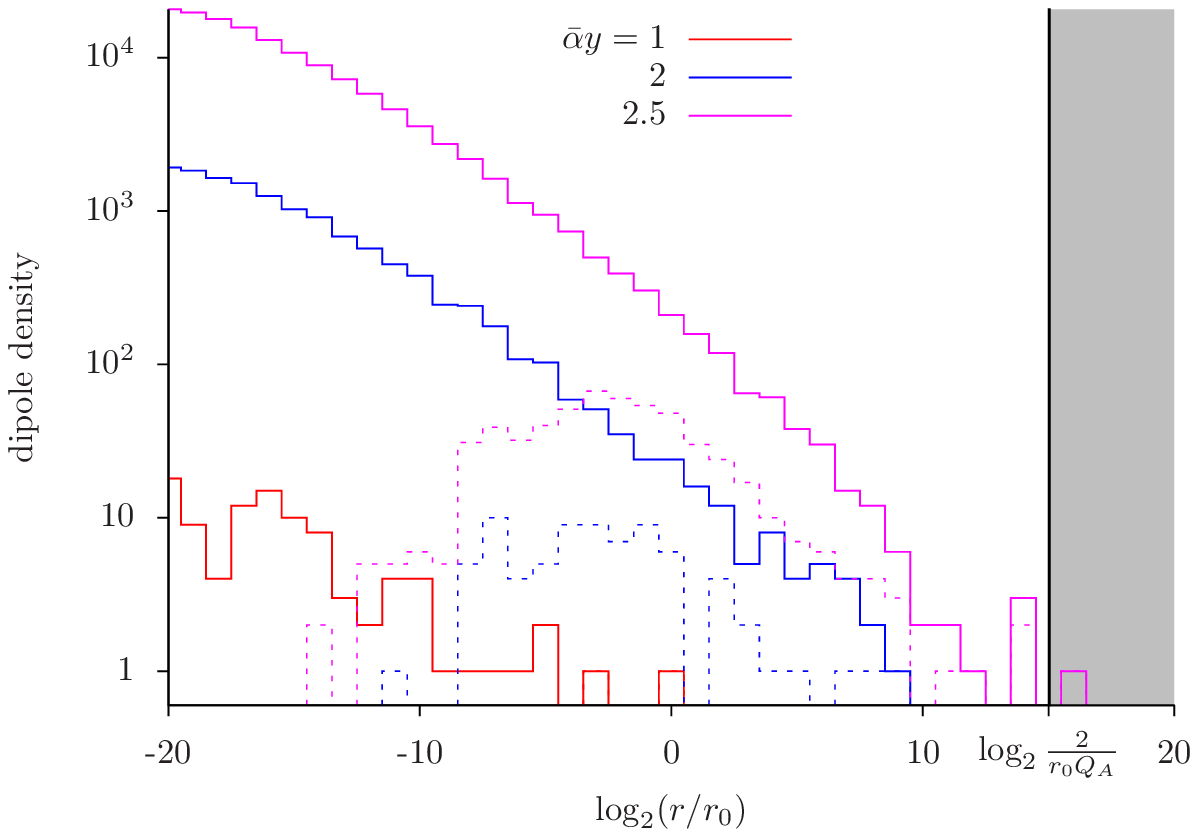}\\
\multicolumn{1}{c}{(a)} & \multicolumn{1}{c}{(b)}
\end{tabular}
\caption{
\label{fig:nucleus}
Two realizations (a) and (b) 
of the same dipole evolution without
saturation in
the simplified model of Ref.~\cite{Munier:2008cg}.
As in Fig.~\ref{fig:dipole}, we plot the
dipole number $n(r_0,r|y)$ in bins of
the logarithm of the dipole size.
The shaded zone represents the target nucleus of saturation
momentum $Q_A$, which absorbs all dipoles of size larger than $2/Q_A$.
Hence this time, the scattering probability in
one realization is 1 if at least one dipole is larger than $2/Q_A$,
and 0 else.
The solid lines represent the dipole
density {\it integrated over the impact parameter}, to which
the nucleus would actually be a priori sensitive, while
the dashed lines represent the density {\it at zero impact
parameter}. We see that the largest dipole is always at zero
impact parameter, which illustrates that it is enough to
understand the properties of the dipole distribution at fixed
impact parameter in order to be able to
compute the scattering amplitude.
The (a)-realization turns out to have 
a larger density than the (b)-realization, 
since the scattering would occur for smaller values of 
$\bar\alpha y$ in the former case than in the latter case.
Note that the scale on the $y$-axis is logarithmic
in both cases.
}
\end{figure}

The way how the scattering probes the quantum evolution is represented in
Figs.~\ref{fig:dipole} and~\ref{fig:nucleus}.
Since the number of dipoles grows exponentially with the rapidity,
$T$ would at some point violate unitarity if the formula for $T_d$
applied without restrictions. But we know that saturation effects (for example
rescatterings, gluon recombination...) may occur in the course of the evolution
so that the effective number of dipoles does never grow
larger than $1/\alpha_s^2$
(see Fig.~\ref{fig:dipole}b).
These saturation effects are crucial in the dipole case, since
the scattering is sensitive to the shape of the bulk
of the distribution of the quantum fluctuations. 
They make the averaging over events nontrivial.\footnote{
In the dipole case, only the front fluctuations have a
significant effect on the amplitude.
This holds true 
for low enough rapidities, namely $\bar\alpha y\ll \ln^3(1/\alpha_s^2)$;
see e.g. \cite{Munier:2009pc} for a discussion 
of how this scale comes about.}\\

In the nucleus case (Fig.~\ref{fig:nucleus}), 
it is the size of the largest dipole
which determines the shape of the amplitude. Therefore, both front and
tip fluctuations must be taken into account. On the other hand, the shape
of the amplitude is less sensitive to saturation effects
in the dipole evolution, at variance
with the dipole case, at least for rapidities parametrically
less than $\frac{1}{\bar\alpha}\ln^3\frac{1}{\alpha_s^2}$.\\

In the regions in which the dipole distribution is deterministic,
the latter essentially has the shape
$\ln \frac{1}{r^2 Q_s^2(y)} \left(r^2 Q_s^2(y)\right)^{-\gamma_0}$ near the
large-size tip of the distribution,
where $Q_s(y)$ is the saturation momentum. Here, $1/Q_s(y)$ may be thought of
as the smallest dipole size $r_0$ for which, in a typical event, there
is an overlap of the dipole number distribution with the target.
The expression of $Q_s(y)$ is given in Ref.~\cite{Mueller:2014fba}.

\section{Phenomenological predictions}

From our model for the fluctuations, an elementary calculation leads to
the shape of the scattering amplitudes in the
dipole-dipole and dipole-nucleus cases:
\begin{equation*}
\left\langle T(r_0|y)\right\rangle
\underset{|r_0| Q_s(y)\ll 1}{\sim}
\left(r_0^2 Q_s^2(y)\right)^{\gamma_0}
\times\begin{cases}
\ln\frac{1}{r_0^2 Q_s^2(y)} & \text{if the target is a nucleus},\\
\ln^2\frac{1}{r_0^2 Q_s^2(y)} & \text{if the target is a dipole}.
\end{cases}
\end{equation*}
The calculation simply averages the amplitude $T^{\text{one dipole}}$
over events, assuming that the stochasticity is fully captured by 
the distribution of the random variables~$\Delta$ and~$\delta$,
see again Ref.~\cite{Mueller:2014fba} for all details and more results.

The fact that the power of the logarithmic prefactor
be different in the dipole-dipole and dipole-nucleus cases
is our main result.


\begin{thebibliography}{10}
\expandafter\ifx\csname natexlab\endcsname\relax\def\natexlab#1{#1}\fi
\providecommand{\enquote}[1]{``#1''}
\expandafter\ifx\csname url\endcsname\relax
  \def\url#1{\texttt{#1}}\fi
\expandafter\ifx\csname urlprefix\endcsname\relax\def\urlprefix{URL }\fi
\providecommand{\eprint}[2][]{\url{#2}}

\bibitem[Mueller and Munier(2014{\natexlab{a}})]{Mueller:2014fba}
A.~Mueller, and S.~Munier, \emph{Phys.Lett.} \textbf{B737}, 303--310
  (2014{\natexlab{a}}), \eprint{1405.3131}.

\bibitem[Munier(2014)]{Munier:2014hga}
S.~Munier, \emph{Proceedings of the PANIC2014 conference}  (2014),
  \eprint{1410.5656}.

\bibitem[Mueller and Munier(2012)]{Mueller:2012bn}
A.~Mueller, and S.~Munier, \emph{Nucl.Phys.} \textbf{A893}, 43--86 (2012),
  \eprint{1206.1333}.

\bibitem[Kovchegov and Levin(2012)]{kovchegov2012quantum}
Y.~Kovchegov, and E.~Levin, \emph{Quantum Chromodynamics at High Energy},
  Cambridge Monographs on Particle Physics, Nuclear Physics and Cosmology,
  Cambridge University Press, 2012, ISBN 9780521112574,
  \urlprefix\url{http://books.google.fr/books?id=f2nHp-NeW6UC}.

\bibitem[Mueller(1994)]{Mueller:1993rr}
A.~H. Mueller, \emph{Nucl.Phys.} \textbf{B415}, 373--385 (1994).

\bibitem[McLerran and Venugopalan(1994)]{McLerran:1993ni}
L.~D. McLerran, and R.~Venugopalan, \emph{Phys.Rev.} \textbf{D49}, 2233--2241
  (1994), \eprint{hep-ph/9309289}.

\bibitem[Mueller and Munier(2014{\natexlab{b}})]{Mueller:2014gpa}
A.~Mueller, and S.~Munier, \emph{Phys.Rev.} \textbf{E90}, 042143
  (2014{\natexlab{b}}), \eprint{1404.5500}.

\bibitem[Salam(1997)]{Salam:1996nb}
G.~Salam, \emph{Comput.Phys.Commun.} \textbf{105}, 62--76 (1997),
  \eprint{hep-ph/9601220}.

\bibitem[Munier et~al.(2008)]{Munier:2008cg}
S.~Munier, G.~Salam, and G.~Soyez, \emph{Phys.Rev.} \textbf{D78}, 054009
  (2008), \eprint{0807.2870}.

\bibitem[Munier(2009)]{Munier:2009pc}
S.~Munier, \emph{Phys.Rept.} \textbf{473}, 1--49 (2009), \eprint{0901.2823}.

\end{thebibliography}
\end{document}